\documentclass[pra,aps,twocolumn,showpacs]{revtex4}
\usepackage{epsfig}
\usepackage{graphicx}
\newcommand{\ket}[1]{|#1\rangle}
\newcommand{\bra}[1]{\langle #1|}

\usepackage{amsthm}
\usepackage{amsfonts}
\newtheorem{Theorem}{Theorem} 
\newtheorem{Lemma}{Lemma}
\newtheorem{Proposition}{Proposition}
\newtheorem{Definition}{Definition}
\begin{document}
\title{Searching for degeneracies of real Hamiltonians using homotopy 
classification of loops in SO($n$)}
\author{Niklas Johansson\footnote{Present address: Department of 
Theoretical Physics, Uppsala University, Box 803, Se-751 08 Uppsala, Sweden} 
and Erik Sj\"oqvist\footnote{Electronic address: eriks@kvac.uu.se}}
\affiliation{Department of Quantum Chemistry, 
Uppsala University, Box 518, Se-751 20 Sweden}
\date{\today}       
\begin{abstract} 
Topological tests to detect degeneracies of Hamiltonians have 
been put forward in the past. Here, we address the applicability 
of a recently proposed test [Phys. Rev. Lett. {\bf 92}, 060406 (2004)] 
for degeneracies of real Hamiltonian matrices. This test relies 
on the existence of nontrivial loops in the space of eigenbases 
SO$(n)$. We develop necessary means to determine the homotopy class 
of a given loop in this space. Furthermore, in cases where the 
dimension of the relevant Hilbert space is large the application 
of the original test may not be immediate. To remedy this deficiency, 
we put forward a condition for when the test is applicable to a 
subspace of Hilbert space. Finally, we demonstrate that applying the 
methodology of [Phys. Rev. Lett. {\bf 92}, 060406 (2004)] to the 
complex Hamiltonian case does not provide any new information.  
\end{abstract}
\maketitle
\section{Introduction}
A particular instance of Berry's discovery \cite{berry84} of geometric
phase factors accompanying adiabatic changes is the occurrence of sign
reversal of eigenfunctions of real Hamiltonians when transported
around certain types of degeneracies. Although already implied in a
work by Darboux \cite{darboux96} in the late 19th century, the 
significance of such sign changes to physics was not realized until 
Longuet-Higgins and coworkers pointed out their existence in molecular 
theory \cite{longuet58,herzberg63}. This latter insight led Mead and 
Truhlar to the notion of the molecular Aharonov-Bohm effect 
\cite{mead79a,mead80}, which has attracted experimental \cite{busch98} 
and theoretical \cite{kendrick97,sjoqvist02} interest recently.

Sign reversal has been noted \cite{arnold78} in characteristic
functions of vibrating membranes whose boundary is changed around
closed paths. Such sign change patterns in the vicinity of
degeneracies have been studied by experiments on microwave resonators
\cite{lauber94,dembowski01} and smectic films \cite{brazovskaia98}. 
The microwave resonator experiments have been interpreted in terms of
both the standard \cite{manolopoulos99} and the off-diagonal 
\cite{pistolesi00} geometric phases, and they have motivated 
further theoretical studies concerning both the geometric phases 
and structure of the wave functions for real Hamiltonians
\cite{manolopoulos99,pistolesi00,samuel01,samuel02}.

It was proved by Longuet-Higgins \cite{longuet75} that sign reversal 
of real electronic eigenfunctions when continuously transported around 
a loop in nuclear configuration space signals the existence of  
degeneracy points inside the loop. This topological result has been 
used to detect conical intersections in LiNaK \cite{varandas79} and 
ozone \cite{xantheas90,ceotto00}. On the other hand, there are 
cases where the Longuet-Higgins test fails, such as, e.g., when 
the loop encircles an even number of conical intersections 
\cite{zwanziger87}. This apparent limitation of the Longuet-Higgins 
theorem was resolved by the present authors \cite{johansson04}, 
who put forward a topological test for degeneracies 
of real matrix Hamiltonians based upon consideration of their
eigenvectors on loops in parameter space. This generalized test was
further proved \cite{johansson04} to be optimal in the sense that it
exhausts all topological information contained in the eigenvectors, 
related to the presence of degeneracies.

Continuous change of the eigenvectors of a real parameter dependent
$n\times n$ matrix Hamiltonian around a closed path in parameter space
may, if all geometric phase factors are unity, be viewed as a loop in
the $n$ dimensional rotation group SO$(n)$. Based upon this
observation, it was proved \cite{johansson04} that if the eigenvectors
correspond to a nontrivial loop in SO$(n)$, then the loop must
encircle at least one point of degeneracy. Thus, in order to apply the
test we need to find a procedure to determine whether the change of
the eigenvectors corresponds to a trivial loop in SO$(n)$ or not. In
Ref. \cite{johansson04}, methods particularly adapted to the special
cases $n=2,3,4$ were presented. The main focus of the present paper is
to provide a method that makes the test applicable to any $n$.

Another important issue for the applicability of the test in Ref. 
\cite{johansson04} arises when noting that $n$ may be very large in 
many realistic scenarios.  For example, the computed electronic
eigenvectors in quantum chemical applications typically live in very
large Hilbert spaces. This could make the test difficult to use in
this important class of problems where degeneracy points play a vital
dynamical role.  To overcome this potential complication, we put
forward a condition for when the test can be applied to subspaces of
the full Hilbert space.

Stone \cite{stone76} demonstrated a topological test that extended
Longuet-Higgins' original test \cite{longuet75} to the complex
Hamiltonian case. In brief, this former test entails that if the
standard geometric phase changes continuously by a nonzero integer
multiple of $2\pi$ for a continuous set of loops in parameter space,
starting and ending with infinitesimally small loops, then this set 
of loops must enclose a degeneracy point. As for Longuet-Higgins' 
test, Stone's test may fail to detect certain kinds of degeneracies. 
This apparently raises the question whether Stone's test can be 
improved. The final concern of this work is exactly to address 
the optimality of Stone's test.

The outline of the paper is as follows. In the next section, we review
the generalized topological test and develop a method to determine
whether loops in SO$(n)$ for arbitrary $n$ are trivial or not. The
main result of this section is contained in Theorem \ref{Thething}.
Sec. \ref{sec:subspaces} contains the condition for the applicability
of the test to subspaces, as summarized in Proposition \ref{subspaces}. 
The optimality of Stone's test for complex Hamiltonians is discussed 
in Sec. \ref{sec:stone}. The paper ends with the conclusions.

\section{Topological test for degeneracies}
\label{sec:test}
Let $H(Q)$ be an $n\times n$ parameter dependent matrix Hamiltonian,
written in the fixed basis $\{ \ket{i} \}_{i=1}^n$ of the $n$
dimensional Hilbert space ${\cal H}$. We suppose that $H(Q)$ is real,
symmetric, and continuous for each $Q = (Q_{1}, \ldots,Q_{d})$ in
parameter space ${\cal Q}$, which we assume to be a simply connected
subset of ${\bf R }^{d}$. Consider a loop $\Gamma$ in ${\cal Q}$. Let
$\{\ket{\psi_{i}(Q)} \}_{i=1}^{n}$ be a positively oriented set of
orthonormalized real eigenvectors of $H(Q)$ for each $Q$ along
$\Gamma$. If there are no points of degeneracy on the loop and if all
the concomitant geometric phase factors of the eigenvectors are unity,
we may define the function $F:\Gamma \rightarrow {\textrm{SO}} (n)$ as
\begin{equation}\label{F}
F(Q) = \left( \begin{array}{ccc} 
\langle 1|\psi_{1}(Q)\rangle & \ldots & \langle 1|\psi_{n}(Q)\rangle \\ 
       \vdots                & \ddots &           \vdots \\
\langle n|\psi_{1}(Q)\rangle & \ldots & \langle n|\psi_{n}(Q)\rangle \\
\end{array} \right) ,  
\end{equation} 
such that $F(\Gamma)$ is a loop in SO$(n)$. Then, the main result 
of Ref. \cite{johansson04} can be summarized in the following 
theorem. 
\begin{Theorem}
If the $n$ eigenvectors of $H(Q)$ represent a nontrivial loop 
$F(\Gamma)$ in SO($n$) when taken continuously around $\Gamma$, 
then there must be at least one degeneracy point of $H(Q)$ on 
every simply connected surface $S$ bounded by $\Gamma$.
\label{maintheorem}
\end{Theorem}

The predictive power of Theorem \ref{maintheorem} stems from the
existence of nontrivial loops in SO$(n\geq 2)$. However, to determine
to which homotopy class a given loop belongs was only treated in the
$n=2,3,4$ cases in Ref. \cite{johansson04}, while the problem to find
such a method for general $n$ was left open. Here, we resolve this
deficiency and put forward an explicit method that treats the general
$n$ case. Let $F:[0,1] \rightarrow \mbox{SO}(n)$ be a loop in
SO$(n)$. Without loss of generality we assume that $F(0) = F(1)$
equals the identity $I$ on SO$(n)$. This can always be achieved by
multiplying the whole loop by $F(0)^{-1} = F(0)^{T}$, an operation
that does not change the homotopy class of $F$.

The space SO($n$) of real orthogonal matrices with unit determinant
forms a Lie group whose corresponding Lie algebra $\mathfrak{so}(n)$ 
is the space of real antisymmetric matrices (see, e.g., Ref. 
\cite{nakahara90}, p. 172). Furthermore, the exponential map  
\begin{eqnarray}
\exp: & \mathfrak{so}(n) & \rightarrow \textrm{SO}(n) 
\nonumber\\
 & A & \mapsto  I + \sum_{k=1}^{\infty}\frac{A^{k}}{k!}
\label{cont}
\end{eqnarray}
is continuous, onto, and $\exp (A)$ is well-defined for all $A \in
\mathfrak{so} (n)$. The function $\log: \textrm{SO}(n) \rightarrow 
\mathfrak{so}(n)$ is defined as the inverse of the exponential map. 
It is multi-valued, i.e., there exist $A \neq B$ such that 
$\exp(A) = \exp(B)$.

By continuity of the exponential map, any curve 
$\widetilde{F}: [0,1]\rightarrow \mathfrak{so}(n)$ satisfying
$\exp(\widetilde{F}(0)) = \exp(\widetilde{F}(1))$ corresponds to a
loop $F$ in SO($n$) via
\begin{equation}\label{widetilde}
F(t) = \exp(\widetilde{F}(t)).
\end{equation}
For brevity we say that continuous curves in $\mathfrak{so}(n)$ for
which $\exp(\widetilde{F}(0)) = \exp(\widetilde{F}(1))$ are {\it
l}-curves. Our objective is to find a correspondence between classes
of {\it l}-curves and the two homotopy classes of loops in SO($n\geq
3$). This will make it possible to deduce whether a loop $F$ in
SO($n$) is trivial by studying its corresponding {\it l}-curve
$\widetilde{F}$. We first make a classification of the {\it l}-curves.
\begin{Definition} 
Let $\widetilde{F}_{0}$ and $\widetilde{F}_{1}$ be l-curves in
$\mathfrak{so}(n)$. If there is a continuous function $L:[0,1]\times
[0,1] \rightarrow \mathfrak{so}(n)$ such that $L(t,0) =
\widetilde{F}_{0}(t)$, $L(t,1) = \widetilde{F}_{1}(t)$, and
$\exp(L(0,s)) = \exp(L(1,s))$ holds for all $s \in [0,1]$, then
$\widetilde{F}_{0}$ and $\widetilde{F}_{1}$ are called
l-homotopic. The function $L$ is called an l-homotopy
between $\widetilde{F}_{0}$ and $\widetilde{F}_{1}$.
\label{l-homotopy}
\end{Definition}
Two {\it l}-curves are thus {\it l}-homotopic if they can be deformed 
into each other through a continuous family consisting solely of 
{\it l}-curves. The following connection between {\it l}-homotopy 
in $\mathfrak{so}(n)$ and homotopy in SO($n$) holds.  
\begin{Proposition} 
If $\widetilde{F}_{0}$ and $\widetilde{F}_{1}$ are {\it l}-homotopic
curves in $\mathfrak{so}(n)$, then their corresponding loops
$F_{0}$ and $F_{1}$ in SO($n$) are homotopic.
\label{homotopy}
\end{Proposition}
\begin{proof}
A homotopy $K:[0,1] \times [0,1] \rightarrow \textrm{SO}(n)$ 
between $F_{0}$ and $F_{1}$ is given by
\begin{equation}
K(t,s) = \exp(L(t,s)),
\end{equation}
where $L$ is as in Definition \ref{l-homotopy}. $K$ defined in this
way is continuous since it is the composition of two continuous
functions. Furthermore $t \mapsto K(t,s)$ is a loop for each $s$ since
$t \mapsto L(t,s)$ is an {\it l}-curve for each $s$.
\end{proof}
Note that we are working with free homotopies in SO($n$) rather than 
based ones. We remark that Proposition \ref{homotopy} implies that 
{\it l}-curves for which $\widetilde{F}(0) = \widetilde{F}(1)$ 
correspond to trivial loops in SO($n$). This fact follows since 
$\mathfrak{so}(n)$ is simply connected. 

The rest of this section is devoted to finding a method to use
Proposition \ref{homotopy} to determine whether a given loop $F$ in
SO$(n)$ is trivial or not. Since we assume that $\exp(\widetilde{F}(0)) =
\exp(\widetilde{F}(1)) = I$, we may choose $\widetilde{F}(0)$
to be the zero matrix. $\widetilde{F}(1)$ is denoted $K$. We show that
an {\it l}-curve $\widetilde{F}$ between the zero matrix and $K$ is
{\it l}-homotopic, either to a point, or to an {\it l}-curve
connecting the zero matrix to a matrix having only one nonzero
$2\times 2$ block given by $2\pi i\sigma_{y}$. In the first case the
corresponding loop $F = \exp(\widetilde{F})$ in SO($n$) is trivial,
and in the second case it is not. We also formulate a simple criterion
that can be used to determine to which ``{\it l}-homotopy class''
$\widetilde{F}$ belongs.

On our way we need the following three lemmas, the first of which is a
standard lemma from matrix theory \cite{axler97}.
\begin{Lemma}\label{antisym}
Let $A$ be an antisymmetric matrix. Then there is an orthogonal matrix
$R$ and a block diagonal matrix $D^{A}$ such that
\begin{equation}
A = RD^{A}R^{T},
\end{equation} 
where the blocks of $D^{A}$ are $2\times 2$ matrices of the form
\begin{equation}\label{Di}
D^{A}_{i} = 
\left( \begin{array}{cc}
0 & \alpha_{i} \\
-\alpha_{i} & 0 \\
\end{array}\right) \equiv i\alpha_{i}\sigma_{y},
\end{equation} 
with $\alpha_{i}\geq 0$. If the dimension $n$ is odd, there is also a
zero $1\times 1$ block. 
\end{Lemma}
For notational convenience, we assume from now on
that $n$ is even. The analysis for odd $n$ is identical.

\begin{Lemma}\label{ort}
Suppose that $\lambda$ is a degenerate eigenvalue of $\exp(A)$, and
that the corresponding eigenspace is $V$. Then $\exp(RAR^{T}) =
\exp(A)$ for any orthogonal matrix $R$ acting nontrivially only on
$V$.
\end{Lemma}
\begin{proof}
\begin{equation}
\exp(RAR^{T}) = R\exp(A)R^{T} = \exp(A),
\end{equation}
where the last equality follows since $\exp(A)$ is $\lambda I$ on $V$.
\end{proof}

\begin{Lemma}\label{tre}
Let $A$ and $B$ be antisymmetric matrices such that $\exp(A) =
\exp(B)$, and let $\widetilde{F}$ be an {\it l}-curve 
between them. Then the following statements hold.
\begin{itemize}
\item[(a)] $\widetilde{F}$ is {\it l}-homotopic to any other 
{\it l}-curve connecting $A$ to $B$. Specifically it is {\it
l}-homotopic to the straight line $t \mapsto (1-t)A + tB$.
\item[(b)] If $X$ is an antisymmetric matrix commuting with both $A$ 
and $B$, then $L(t,s) = \widetilde{F}(t) - sX$ is an {\it l}-homotopy. 
The {\it l}-curve $L(t,1)$ connects $A-X$ to $B-X$.
\item[(c)] Let $\lambda$ be a degenerate eigenvalue of $\exp(A)$, and $V$ 
be the corresponding eigenspace. If $R$ is an orthogonal transformation 
with unit determinant acting nontrivially only on $V$, then 
$\widetilde{F}$ is {\it l}-homotopic to an {\it l}-curve connecting 
$RAR^{T}$ to $B$.
\end{itemize}
\end{Lemma}
\begin{proof}
(a) Follows since $\mathfrak{so}(n)$ is a vector space, and thus
simply connected. To prove (b) we note that $L$ is continuous and that
\begin{eqnarray}
\exp(L(0,s)) & = & \exp(A-sX) = \exp(A)\exp(-sX) 
\nonumber \\ 
 & = & \exp(B)\exp(-sX) 
\nonumber \\ 
 & = & \exp(B-sX) = \exp(L(1,s)) .
\end{eqnarray}
For (c), note that since $R$ acts nontrivially only on $V$, it is
possible to write $R = \exp(C)$, where $C$ is an antisymmetric matrix
whose null space contains the orthogonal complement of $V$. This means
that $\exp(r C)$ acts nontrivially only on $V$ for any real number 
$r$. Lemma \ref{ort} is thus applicable, and $\exp[\exp(-r
C)A\exp(r C)] = \exp(A)$ for any $r$. We may thus define the {\it
l}-homotopy
\begin{equation}
L(t,s) = \exp((1-t)sC)\widetilde{F}(t)\exp((t-1)sC).
\end{equation}
We see that $L(t,0) = \widetilde{F}(t)$, $L(0,1) = RAR^{T}$, and
$L(1,1) = B$ as required.
\end{proof}
Before we proceed it is convenient to introduce some notation to
describe block diagonal matrices. First, $A_{1}\oplus \ldots \oplus
A_{m}$ will denote a block diagonal matrix with blocks
$A_{1},\ldots,A_{m}$. Secondly, for any numbers $\lambda_{1}, \ldots
,\lambda_{n/2}$ we define $[\lambda_{1},\ldots,\lambda_{n/2}] \equiv
(i \lambda_{1}\sigma_{y}) \oplus \ldots \oplus (i \lambda_{n/2}
\sigma_{y})$. Note that any two matrices of this form
commute. Thirdly, $I_{m}$ and $0_{m}$ denote the $m$-dimensional unit
and zero matrix, respectively.

Now, we return to our {\it l}-curve $\widetilde{F}$ connecting $0_{n}
= \widetilde{F}(0)$ to $K \equiv \widetilde{F}(1)$. Let $R$ be as in
Lemma \ref{antisym}, so that $K = RD^{K}R^{T}$, with $D^{K}=[
\alpha_{1}, \ldots,
\alpha_{n/2}]$. Note that since $\exp(K) = I$, we must have
$\alpha_{i} = 2\pi k_{i}$ for some integers $k_{i}$. At the possible
cost of having some $k_{i} < 0$ we may assume $R$ to have unit
determinant. This implies that $\widetilde{F}$ is {\it l}-homotopic to
any {\it l}-curve connecting $0_{n}$ and $D^{K}$. To see this, let $R
= \exp(C)$ and define
\begin{equation}
L(t,s) = \exp(-sC)\widetilde{F}(t)\exp(sC).
\end{equation}
Clearly $L(t,0) = \widetilde{F}(t)$ and $L(t,1)$ goes from $0_{n}$ to
$D^{K}$. Furthermore, $L$ is continuous, and
\begin{equation}
\exp(L(0,s)) = I = \exp(L(1,s)). 
\end{equation}
Thus, $L$ is an {\it l}-homotopy, and by Lemma \ref{tre}(a), 
$\widetilde{F}$ is {\it l}-homotopic to the {\it l}-curve
\begin{equation}\label{block}
\widetilde{F}_{block}(t) = tD^{K}.
\end{equation}
Our goal is to make as many as possible of the $k_{i}$ disappear
through {\it l}-homotopies. We begin by reducing each of them to zero 
or one. Assume that $k_{i} \geq 2$. The case $k_{i} \leq -1$ can be
treated similarly. Define $Y_{i} = [0,\ldots,0,2\pi,0,\ldots,0]$, where
the $2\pi$ appears at the $i$th place. $Y_{i}$ commutes with $0_{n}$
and $D^{K}$, i.e.,
\begin{equation}
L(t,s) = \widetilde{F}_{block}(t) - sY_{i},
\end{equation}
is an {\it l}-homotopy transforming $\widetilde{F}_{block}$ into
$\widetilde{F}_{i} = (t-1)Y_{i}+t(D^{K}-Y_{i})$. This {\it l}-curve starts
at $-Y_{i}$ and terminates in the matrix
$D^{K}-Y_{i}=2\pi[k_{1},\ldots,k_{i-1},
k_{i}-1,k_{i+1},\ldots,k_{m}]$. Furthermore $\exp(-Y_{i}) = I$. This
makes Lemma \ref{tre}(c) applicable, $V$ being the whole space. For 
$i\geq 2$ (the case $i=1$ is similar), we choose the orthogonal 
transformation $R$ as 
\begin{equation}
R = I_{2i-3}\oplus (-I_{1})\oplus \sigma_{x} \oplus I_{n-2i}, 
\end{equation}
where $\sigma_{x}$ is the $x$ component of the standard Pauli matrices. 
$R$ thus defined has unit determinant, and 
\begin{eqnarray}
R(-Y_{i})R^{T} & = & 0_{2i-2}\oplus(-i2\pi\sigma_{x}\sigma_{y}\sigma_{x})
\oplus 0_{n-2i}
\nonumber \\ 
 & = & 0_{2i-2}\oplus(i2\pi\sigma_{y})\oplus 0_{n-2i} = Y_{i}.
\end{eqnarray}
Consequently, by Lemma \ref{tre}(c) $\widetilde{F}_{i}$ is 
{\it l}-homotopic to an {\it l}-curve between $Y_{i}$ and $D^{K}-Y_{i}$, 
and thus to the {\it l}-curve
\begin{equation}
\widetilde{F}_{block,i-}(t) = t(D^{K}-2Y_{i}).
\end{equation}
If we compare this to Eq. (\ref{block}), and note that
$D^{K}-2Y_{i}=2\pi[k_{1},\ldots, k_{i-1}, k_{i}-2, k_{i+1}, \ldots,
k_{n}]$, we see that we have reduced $k_{i}$ by two.

Proceeding in this way we may show that $\widetilde{F}_{block}$ is
{\it l}-homotopic to $\widetilde{F}_{red}(t) = tP$, where $P = 2\pi
[\delta_{1}, \ldots, \delta_{n}]$, and $\delta_{i}$ is defined by
\begin{equation}\label{delta}
\delta_{i} = 
\left\{ \begin{array}{ll} 1 & \textrm{if} \ k_{i} \ \textrm{is odd} \\
0 & \textrm{if} \ k_{i} \ \textrm{is even.} \end{array} \right.
\end{equation}
Our next task is to reduce the number of nonzero $\delta_{i}$ to one
or zero. We will show that any pair $\delta_{i} = \delta_{j} = 1$ can
be eliminated through {\it l}-homotopies. The procedure for doing this 
is similar to the reduction of the $k_{i}$. 

Suppose that $i<j$. First deform $\widetilde{F}_{red}$ by
\begin{equation}
L(t,s) = \widetilde{F}_{red}(t) - \frac{s}{2}(Y_{i}+Y_{j}),
\end{equation}
yielding $\widetilde{F}_{i,j}(t)$ starting at $-Y \equiv
-\frac{1}{2}(Y_{i}+Y_{j})$. Note that $\exp(-Y)$ is four-fold
degenerate with eigenvalue $-1$. The eigenspace is
$\textrm{supp}(Y)$. We apply Lemma \ref{tre}(c) with $R$ defined by
\begin{equation}
R = I_{2(i-1)}\oplus \sigma_{x} \oplus I_{2(j-i-1)} \oplus 
\sigma_{x} \oplus I_{n-2j}.
\end{equation}
This matrix is orthogonal and has unit determinant. Also, we may 
verify that
\begin{equation}
R(-Y)R^{T} = Y.
\end{equation}
Consequently, as we went from $\widetilde{F}_{block}(t)=tD^{K}$ to
$\widetilde{F}_{block,i-}(t) = t(D^{K}-2Y_{i})$, we can go from
$\widetilde{F}_{red}(t) = tP$ to $\widetilde{F}_{red, -i, -j}(t) =
t(P-2Y) = t(P-Y_{i}-Y_{j})$. The matrix $P-Y_{i}-Y_{j}$ has the same
structure as $P$, but has $\delta_{i} = \delta_{j} = 0$.  Continuing
in this fashion we can reduce the number of nonzero $\delta_{i}$ to
one (zero) if this number was odd (even) to begin with. Note that this
number is odd (even) exactly when $\sum_{i=1}^{n/2} k_{i}$ is odd
(even). At long last we arrive at the following main result. 
\begin{Theorem}\label{Thething}
Suppose that $F$ is a loop in SO($n$) starting at $I$ and that
$\widetilde{F}$ is an {\it l}-curve that maps to $F$ under the
exponential map Eq. (\ref{widetilde}). Then there is an orthogonal
transformation $R$ so that
\begin{equation}
\widetilde{F}(1) = R 2\pi (ik_1\sigma_y) \oplus 
\ldots \oplus (ik_{n/2}\sigma_y)R^{T},
\end{equation}
for some integers $k_{1},\ldots,k_{n/2}$. Furthermore $F$ is
trivial if and only if $h \equiv \sum_{i=1}^{n/2} k_{i}$ is even.
\end{Theorem}
Note that in case of odd $n$, the matrix $R^T \widetilde{F}(1) R$ has
one zero $1\times 1$ block that can be ignored.
\begin{proof}
The theorem follows from the above discussion, and from Proposition
\ref{homotopy}. If $h$ is even, then $\widetilde{F}$ is
{\it l}-homotopic to a point. Otherwise it is {\it l}-homotopic to
$\widetilde{F}_{Y}(t) = tY_{i}$ for some $i$, which makes  
$\exp(\widetilde{F}_{Y}(t))$ nontrivial.
\end{proof}
Theorem \ref{Thething} reduces the task of determining the homotopy
class of a loop in SO$(n)$ that starts at the identity to computing
the logarithm $\widetilde{F}$ \cite{gallier03} and block diagonalizing
its ending point.

We illustrate the procedure by determining the homotopy class of a
loop in SO(3). Let 
\begin{equation}\label{FTxtNummer2}
F(\theta) = \left( \begin{array}{ccc}
\frac{1}{2}(\cos\theta + 1) & -\frac{1}{\sqrt{2}}\sin\theta  & 
\frac{1}{2}(\cos\theta-1)\\
\frac{1}{\sqrt{2}}\sin\theta& \cos\theta & \frac{1}{\sqrt{2}}\sin\theta\\
    \frac{1}{2}(\cos \theta -1)        &  
-\frac{1}{\sqrt{2}}\sin\theta       &     \frac{1}{2}(\cos\theta + 1)\\
\end{array}\right),
\end{equation}
where $\theta \in [0,2\pi]$ parametrizes the loop. This loop appears
in the analysis \cite{johansson04} of the $T\otimes \tau_{2}$
Jahn-Teller system \cite{obrien89}. In fact, Eq. (\ref{FTxtNummer2})
represents the loop in Eq. (4) of Ref. \cite{johansson04} multiplied
by its inverse at $\theta = 0$, so that $F(0) = I$. For each $\theta$,
$F(\theta)$ is a three-dimensional rotation whose angle $\phi$ and
axis ${\bf \hat{v}}$ of rotation are given by
\begin{eqnarray}\label{angleandaxis}
\phi(\theta) &=& \theta,\nonumber\\
{\bf \hat{v}}(\theta) & = & (\hat{v}_1(\theta), \hat{v}_2(\theta), 
\hat{v}_3(\theta)) = \frac{1}{\sqrt{2}}(-1,0,1) \label{vNummer2}.
\end{eqnarray}
Note that ${\bf \hat{v}}$ is undefined at $\theta = 0$ and
$2\pi$, since $F(0) = F(2\pi) = I$. It is straightforward to check that
$\exp(\widetilde{F}(\theta)) = F(\theta)$ if we define
\begin{eqnarray}
\widetilde{F}(\theta) & = & \phi(\theta)
\left( \begin{array}{ccc}   0   & -\hat{v}_{3}(\theta)  & 
\hat{v}_{2}(\theta)  \\ 
                         \hat{v}_{3}(\theta) &   0    &  
-\hat{v}_{1}(\theta)  \\ 
                          - \hat{v}_{2}(\theta) &  
\hat{v}_{1}(\theta) &   0     \\ 
\end{array} \right)
\nonumber \\ 
 & = & \frac{\theta}{\sqrt{2}} 
\left( \begin{array}{ccc}   0   & -1 & 0\\ 
                         1 & 0 & 1\\ 0& -1& 0 \\
\end{array} \right).
\end{eqnarray}
This curve is continuous everywhere, starts at the zero matrix and
ends, by continuity, at
\begin{eqnarray}
K = \pi \sqrt{2} \left( \begin{array}{ccc}   0   & -1 & 0\\ 
                         1 &   0    &  1\\ 
                          0&  -1&   0     \\ 
\end{array} \right).
\end{eqnarray}
The orthogonal transformation 
\begin{eqnarray}
R = \frac{1}{\sqrt{2}}
\left( \begin{array}{ccc}   -1 &  0 & 1\\ 
                             0&   \sqrt{2}    &  0 \\ 
                          1& 0 & 1\\ 
\end{array} \right) 
\end{eqnarray}
block diagonalizes $K$. Explicitly, we have 
\begin{equation}
D^{K} = R^{T}KR = \left( \begin{array}{ccc}   0 &  0 & 0\\ 
                               0 &  0    &  2\pi \\ 
                               0 &  -2\pi  &  0\\ 
\end{array} \right).
\end{equation}
Thus, $h=k_{1}=1$ and the loop is nontrivial.

\section{Application to subspaces}
\label{sec:subspaces}
Here, we demonstrate how to apply the test to a subspace of Hilbert 
space. Specifically, we show that if $p<\dim \mathcal{H}$ eigenvectors 
of a parameter dependent Hamiltonian can be well approximated by 
their projections in some fixed $p$ dimensional subspace of Hilbert 
space, then the $p$ dimensional version of the test in Ref. 
\cite{johansson04} is applicable. This result may be of use in 
quantum chemical applications, where the detection of degeneracy 
points for electronic Hamiltonian matrices of large dimension becomes 
pertinent. 

Suppose that $\{|i\rangle\}_{i=1}^{p}$ is an orthonormal set of
fixed real vectors. We consider $p$ eigenvectors
$\{|\psi_{i}(Q)\rangle \}_{i=1}^{p}$ of $H(Q)$ and their projections
in ${\textrm{Span}} \{|i \rangle\}$. Define
\begin{eqnarray}
P & = & \sum_{i=1}^{p}|i\rangle \langle i|, 
\nonumber \\
\ket{\phi_{i}(Q)} & = & P\ket{\psi_{i}(Q)}, 
\nonumber \\ 
\ket{\phi_{i}^{\perp} (Q)} & = & \ket{\psi_{i}(Q)} - 
\ket{\phi_{i}(Q)} 
\label{phi}
\end{eqnarray}
The eigenvectors $\ket{\psi_{i}(Q)}$ are mutually orthogonal,
but $\ket{\phi_{i}(Q)}$ need not be. In fact they may even be
linearly dependent. This, however, can occur only if $\bra{\phi_{i}(Q)} 
\phi_{i}(Q)\rangle$ is sufficiently small for some $i$. This is 
formalized in Proposition \ref{lid}. For the proof we need
the following. 
\begin{Lemma}\label{simplex}
Let $v_1,\ldots,v_p$ be linearly dependent unit vectors 
in a real vector space. Then for some pair $v_{j}, v_{k}$ 
with $j\neq k$ we have 
\begin{equation}
|v_{j} \cdot v_{k}| \geq \frac{1}{p-1}.
\end{equation}
\end{Lemma}
\begin{proof}
Let ${\textrm{Span}} \{ v_1,\ldots,v_p \}$ have dimension $m$. 
It is enough to prove the statement for $m=p-1$, since if it is
false for some $m$, then it is false for all higher $m$.

Intuitively, to make all scalar products as small as possible, we need
to ``spread'' the vectors as much as possible, i.e., the vectors
should point to the vertices of a regular $(p-1)$-simplex
\cite{remark1}. The scalar product between any two distinct vectors is
then $-\frac{1}{p-1}$ \cite{parks02}. This proves the statement.
\end{proof}
\begin{Proposition}\label{lid}
Suppose that 
\begin{equation}\label{req} 
\langle \phi_{i}| \phi_{i}\rangle > 1-\frac{1}{p}
\end{equation}
holds for all $i$. Then $\{|\phi_{i}\rangle \}$ is a linearly
independent set.
\end{Proposition}
\begin{proof}
Assume that Eq. (\ref{req}) holds for each $i$, and that the vectors
$|\phi_{i}\rangle$ are linearly dependent. We show that this leads to
a contradiction. With $\ket{\phi_i} = \ket{\phi^{N}_i} \sqrt{\bra{\phi_i} 
\phi_{i}\rangle}$, there are by Lemma \ref{simplex} $j \neq k$ such that
\begin{eqnarray}
|\langle \phi_{j}| \phi_{k}\rangle| & = & 
\sqrt{\langle \phi_{j}|\phi_{j}\rangle} 
\sqrt{\langle \phi_{k}| \phi_{k}\rangle} 
|\langle \phi^{N}_{j}| \phi^{N}_{k}\rangle| 
\nonumber \\ 
 & > & \left(1-\frac{1}{p}\right) \frac{1}{p-1} = \frac{1}{p}.
\label{great}
\end{eqnarray}
Note however that
\begin{equation}
0 = \langle \psi_{j}| \psi_{k}\rangle = \langle \phi_{j}|
\phi_{k}\rangle + \langle \phi^{\perp}_{j}| \phi^{\perp}_{k}\rangle
\end{equation}
and thus that
\begin{eqnarray}
|\langle \phi_{j}| \phi_{k}\rangle| & = & |\langle \phi^{\perp}_{j}|
\phi^{\perp}_{k}\rangle| \leq \big\{ \langle \phi^{\perp}_{j}|
\phi^{\perp}_{j}\rangle \langle \phi^{\perp}_{k}|
\phi^{\perp}_{k}\rangle \big\}^{1/2} 
\nonumber \\
 & = & \big\{(1-\langle \phi_{j}| \phi_{j}\rangle) 
(1-\langle \phi_{k}| \phi_{k}\rangle) \big\}^{1/2} 
\nonumber \\ 
 & < & \frac{1}{p} ,
\end{eqnarray}
contradicting Eq. (\ref{great}).
\end{proof}
We are now in a position to give a condition for when the
$p$ dimensional test is applicable to a subspace of Hilbert space.
Let $S$ be a simply connected surface in $\cal Q$, bounded by the loop
$\Gamma$, and let the $p$ eigenvectors along $\Gamma$ be
denoted $\{|\psi_{i}(Q)\rangle \}_{i=1}^{p}$. Assume that for
$|\phi_{i}(Q)\rangle$ defined by Eq. (\ref{phi}), the inequality
\begin{equation}\label{ineq}
\langle \phi_{i}(Q)|\phi_{i}(Q)\rangle > 1-\frac{1}{p},
\end{equation}
holds for each $i$, and for each $Q \in S$. The set $\{|\phi_{i}
(Q)\rangle \}_{i=1}^{p}$ is then linearly independent by Proposition
\ref{lid}. This means that the Gram-Schmidt orthonormalization
procedure can be applied to $\{|\phi_{i} (Q)\rangle \}_{i=1}^{p}$,
for any $Q$ in $S$. This procedure is continuous, and produces an
orthonormal set $\{ |\phi^{GS}_{i}(Q)\rangle \}_{i=1}^{p}$, which can
be interpreted as an element $F(Q) \in \textrm{SO}(p)$. Thus, given that
Eq. (\ref{ineq}) holds and that $H(Q)$ is nondegenerate on $S$, we
have a continuous function $F:S\rightarrow \textrm{SO}(p)$. By the same
reasoning as in Ref. \cite{johansson04} we arrive at the following 
result.    
\begin{Proposition}
Suppose that the projections $|\phi_{i}(Q)\rangle$ of the $p$ 
eigenvectors of $H(Q)$ satisfy Eq. (\ref{ineq}) for each $Q\in S$. 
Suppose furthermore that $F(Q)$ defined as above traces out a
nontrivial loop in SO($p$) as $Q$ varies along $\Gamma$. Then $H(Q)$
becomes degenerate somewhere in $S$.
\label{subspaces}
\end{Proposition}
Note that the result in Proposition \ref{subspaces} concerns the 
exact Hamiltonian $H(Q)$, but is based upon the behavior of the 
approximate eigenvectors $\ket{\phi_i(Q)}$. 

\section{Stone's test}
\label{sec:stone} 
The original test by Longuet-Higgins \cite{longuet75}, as well as its
generalization \cite{johansson04}, suffers from the limitation of
being applicable only to real Hamiltonians. When the test of
Longuet-Higgins is applicable, the loop in parameter space maps to an
open curve in ${\bf R}^{n}$ representing the Hilbert space. In this
case the corresponding loop in the space of states ${\bf RP}^{n-1}$ is
nontrivial. Similarly, the generalization makes use of the existence
of nontrivial loops in SO$(n)$, the space of eigenbases. A pertinent
question is whether there exist analogous results for the general
complex case.

For a generic Hamiltonian, the degenerate subsets of parameter space
have co-dimension $3$ \cite{mead79b,stone76}, meaning that any such
test must consider eigenvectors on a closed surface, rather than on a
closed loop. An eigenstate taken around the surface represents a
$2$-loop in projective Hilbert space, being the $n-1$ dimensional
complex projective space ${\bf CP}^{n-1}$.

Stone \cite{stone76} put forward a topological test relating the
behavior of a single eigenvector on a closed surface $S$ in parameter
space, to the presence of degeneracies inside the surface. Potentially, 
this test might be possible to generalize in the same manner as the 
one by Longuet-Higgins. However, as shown below, this is impossible.

Despite that it preceded Berry's work \cite{berry84} by eight years,
Stone's test is conveniently formulated in the language of geometric
phases. The surface $S$ is swept out by a continuous set
$\{L_{i}\}_{i=1}^{N}$ of loops, where $L_{1}$ and $L_{N}$ are
infinitesimally small. The cyclic geometric phases
$\{\gamma_{i}\}_{i=1}^{N}$ along these loops are modulo $2\pi$
quantities. However, if we require continuity in the index $i$ and
choose $\gamma_{1} = 0$, $\gamma_{N}$ becomes uniquely determined, and
equal to $2\pi k$ for some integer $k$. Stone proved that if $k \neq
0$, then there must be a degeneracy point somewhere inside $S$. The
integer $k$ can be topologically interpreted as labeling the homotopy
class to which the 2-loop represented by the states around $S$
belongs. Equivalently, $k$ characterizes the topological structure of
the monopole bundle with fiber U$(1)$ representing state vectors and
base space $S^{2}$ representing the surface $S$ (see, e.g.,
Ref. \cite{nakahara90}, p. 320).

It turns out that it is possible to define a global and continuous
state vector around $S$ if and only if $k=0$, i.e., if and only if
Stone's test does not signal a degeneracy. Let us try to construct a
test that works even for some cases when $k=0$ by considering a
complete set of eigenvectors around $S$. We then have a continuous
function from the surface $S$ to the space of eigenbases U($n$). This
can contain topological information only if U($n$) contains nontrivial
$2$-loops. This, however, is not the case (see, e.g.,
Ref. \cite{nakahara90}, pp. 120-121). Consequently, Stone's test
exhausts all topological information contained in the eigenvectors
around $S$.

We conclude this section by noting that, while Stone's test is optimal 
for general Hamiltonians, topological tests similar to that of Ref. 
\cite{johansson04} may be constructed for Hamiltonians obeying 
additional symmetries.  

\section{Conclusions} 
The need to demonstrate whether or not there exist degeneracy points 
in the spectra of Hamiltonians is of relevance in many fields of 
physics. For example, such points are abundant in molecular systems
\cite{truhlar03} and are important because they signal a breakdown of 
the Born-Oppenheimer approximation. Another instance where the
presence of degeneracy points become pertinent is in the recently
proposed paradigm of adiabatic quantum computation
\cite{farhi00,farhi01}, whose efficiency relies crucially upon 
the presence of nonvanishing energy gaps along certain paths in
parameter space. 

Topological tests to detect degeneracies have been put forward in the
past by Longuet-Higgins \cite{longuet75} and Stone \cite{stone76}.
More recently, the present authors \cite{johansson04} extended
Longuet-Higgins' test by consideration of complete sets of
eigenvectors of real parameter dependent Hamiltonian matrices as paths
in SO$(n)$. This extended test can detect degeneracies even in cases
where Longuet-Higgins' original test fails.

In this paper, we have put forward a method that makes the topological
test in Ref. \cite{johansson04} applicable to any dimension $n$ of the
Hamiltonian matrix. This method is based upon the multi-valuedness of
the function $\log : \textrm{SO}(n) \mapsto \mathfrak{so}(n)$, that
connects SO$(n)$ with its corresponding Lie algebra $\mathfrak{so}(n)$
of real antisymmetric matrices. We have further demonstrated under
what conditions the topological test in Ref. \cite{johansson04} is
applicable to subspaces of the full Hilbert space. These two major
findings of the present paper open up the possibility to use the test
in realistic scenarios, such as, e.g., for computed electronic
eigenvectors in various molecular systems or for the eigenfunctions 
of quantum billiards.  Our final result concerns Stone's test for 
degeneracies of complex Hamiltonians. We have shown that Stone's test
is optimal in the sense that no other topological test can do better 
in detecting degeneracies in systems that need a description in terms 
of general complex Hamiltonians.

\section*{Acknowledgments} 
We wish to thank David Kult and Johan {\AA}berg for discussions and 
useful comments.


\begin{thebibliography}{99}
\bibitem{remark1} A regular $n$-simplex is the generalization to 
higher dimensions of an equilateral triangle (the $2$-simplex) 
and a tetrahedron (the $3$-simplex).
\end{thebibliography}
\end{document}